\documentclass{mixdes}
\usepackage[utf8]{inputenc}
\usepackage{graphicx,amsmath,amsfonts}
\usepackage{courier}
\usepackage[english]{babel}
\usepackage{scstuff}
\usepackage[mediumspace,mediumqspace,squaren,cdot]{scSIunits}
\usepackage[LGR,T1]{autofe}
\usepackage{cite}
\usepackage[final]{microtype}
\usepackage[svgnames]{xcolor}
\usepackage[pdftex, unicode, bookmarks=false, bookmarksnumbered=true,
  pdfpagemode={UseOutlines}, plainpages=false, pdfpagelabels=true,
  colorlinks=true, linkcolor={DarkBlue}, citecolor={DarkBlue}, 
  urlcolor={DarkBlue}]{hyperref}

\newcommand{\mailto}[1]{\href{mailto:#1}{\nolinkurl{#1}}}
\newcommand{\doi}[1]{\href{http://dx.doi.org/#1}{\nolinkurl{#1}}}

\renewcommand\normalsize{\fontsize{10.1}{12.5}\selectfont}

\title{%
  Interleaving techniques\\
  for high-throughput chaotic noise generation in CMOS}

\author{%
  \textbf{S. Callegari, R. Rovatti}\\
  D.E.I.S. - Università di Bologna, ITALY
  \\[1ex]
  \textbf{G. Setti}\\
  D.I. - Università di Ferrara, ITALY%
  \thanks{This is a pre-print version of a
    paper presented at the 7th International Conference Mixed Design of
    Integrated Circuits and Systems (MIXDES 2000).  Published paper available
    in ``Proceedings of the 7th International Conference Mixed Design of
    Integrated Circuits and Systems (MIXDES 2000)''. Cite as:\protect\\[1ex]
    S.~Callegari, R.~Rovatti, G.~Setti, `` Interleaving techniques for
    high-throughput chaotic noise generation in CMOS'', in Proc. of MIXDES 200,
    Gdynia, Poland, Jun.~2000, pp.~139–144.}%
}

\hypersetup{%
  pdftitle={%
    Interleaving techniques for high-throughput chaotic noise generation in
    CMOS},
  pdfauthor={%
    Sergio Callegari,
    Riccardo Rovatti,
    Gianluca Setti}}


\date{}
  
\graphicspath{{./Figures/}{./}}

\addunit{\sample}{sample}
\addunit{\decibel}{dB}

\theAbstract{%
  An interleaving technique is proposed to enhance the throughput of
  current-mode CMOS discrete time chaotic sources based on the
  iteration of unidimensional maps. A discussion of the reasons and
  the advantages offered by the approach is provided, together with
  analytical results about the conservation of some major statistical
  features. As an example, application to an FM-DCSK communication
  system is proposed. To conclude, a sample circuit capable of
  \Unit{20}{\mega\sample\per\second} is presented.}

\begin{document}

\maketitle

\section*{Introduction}
\begin{figure}[ht]
  \centering
  \resizebox{0.8\linewidth}{!}{\includegraphics{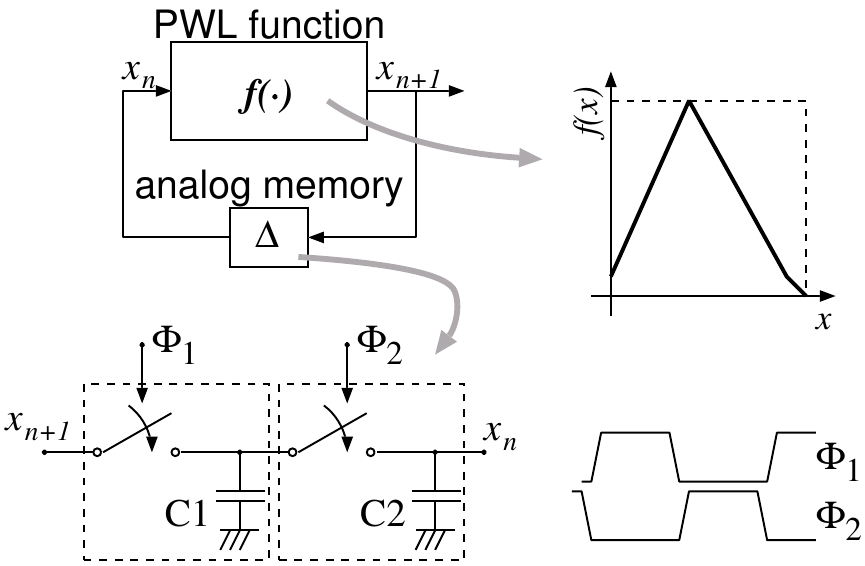}}
  \caption{\label{fig:TTMchaos}%
    Chaos based noise generation, exploiting a discrete time one
    dimensional system. The circuit implements $x_{n+1}=f(x_n)$, where
    $f(\cdot)$ is a suitable non-invertible function.}
\end{figure}

Interest in the hardware implementation of chaotic systems has
recently been boosted by the expectations in fields such as
communication systems \cite{Kennedy:ECCTD99}, biologically-inspired
computation \cite{Clarkson:NNW-1993-5}, noise generation, EMI
reduction \cite{Rovatti:ECCTD99}, etc.  Many applications depend on
the existence of pseudo-random data streams of given statistical
properties and can directly benefit from chaos-based generators. In
fact, the collocation of chaos at the borderline between randomness
and causality can be deployed for building extremely simple analog
CMOS noise-like sources characterized by small areas and power
requirements \cite{Delgado:EL-1993-25,Callegari:ISCAS97}.
These are commonly based on discrete time models like the one in
Figure~\ref{fig:TTMchaos} \cite{Delgado:EL-1993-25,Devaney:CDS-1989}.
Note that for the implementation of the individual building blocks,
the current mode approach is currently the best established one.
Regrettably, the systems proposed so far exhibit a limited data rate
which hinders their applicability. Even with unconventional design
optimizations \cite{Callegari:ECCTD99}, it is difficult to rise
data rates over a few \mega\hertz.

Herein we propose the addressing of throughput issues by means of
hardware resources replication and parallel operation. Particularly,
we propose a technique for order-2 concurrency which allows two
identical independent chaotic sources to share a large amount of their
hardware.  The approach is validated by: \emph{i.} analytical results
about major output statistical properties; \emph{ii.} an application
example to an FM-DCSK communication system \cite{Kennedy:ECCTD99} and
\emph{iii.} the simulation of a sample circuit designed over a
conventional \Unit{0.8}{\micro\metre} CMOS technology and capable of
\Unit{20}{\mega\sample\per\second}.

\section*{Analog register options}
With reference to the model in Figure~\ref{fig:TTMchaos} and to
operating frequencies, one of the most critical sections is the memory
which is required for keeping $x_n$ stable while the map circuit
evaluates $x_{n+1}$. Analog operation is a pre-requisite for truly
chaotic behaviour, yet this introduces errors in the in-loop signal
path which must be kept extremely low as they have an immediate impact
on the output statistics of the chaotic source
\cite{Callegari:NOLTA98}. As expectable, speed-accuracy trade-off do
normally exist.

Since the analog memory synchronizes the circuit to an
external clock and is never allowed to provide a \emph{transparent}
connection from input to output, its operation is actually as the
analog counterpart of a digital register. Just as a digital register,
it can be built up of more elementary \emph{latching elements}, in
this case sample-and-hold (SH) or track-and-hold (TH) units.
Hence, three main degrees of freedom are allowed to the designer:
choice of the register architecture, the choice in between TH or SH
elements, and the choice of the particular circuit for the SH\emph{s} or
TH\emph{s}. Obviously, the three choices are interrelated.

\begin{figure}[ht]
  \centering\includegraphics[width=0.95\lw]{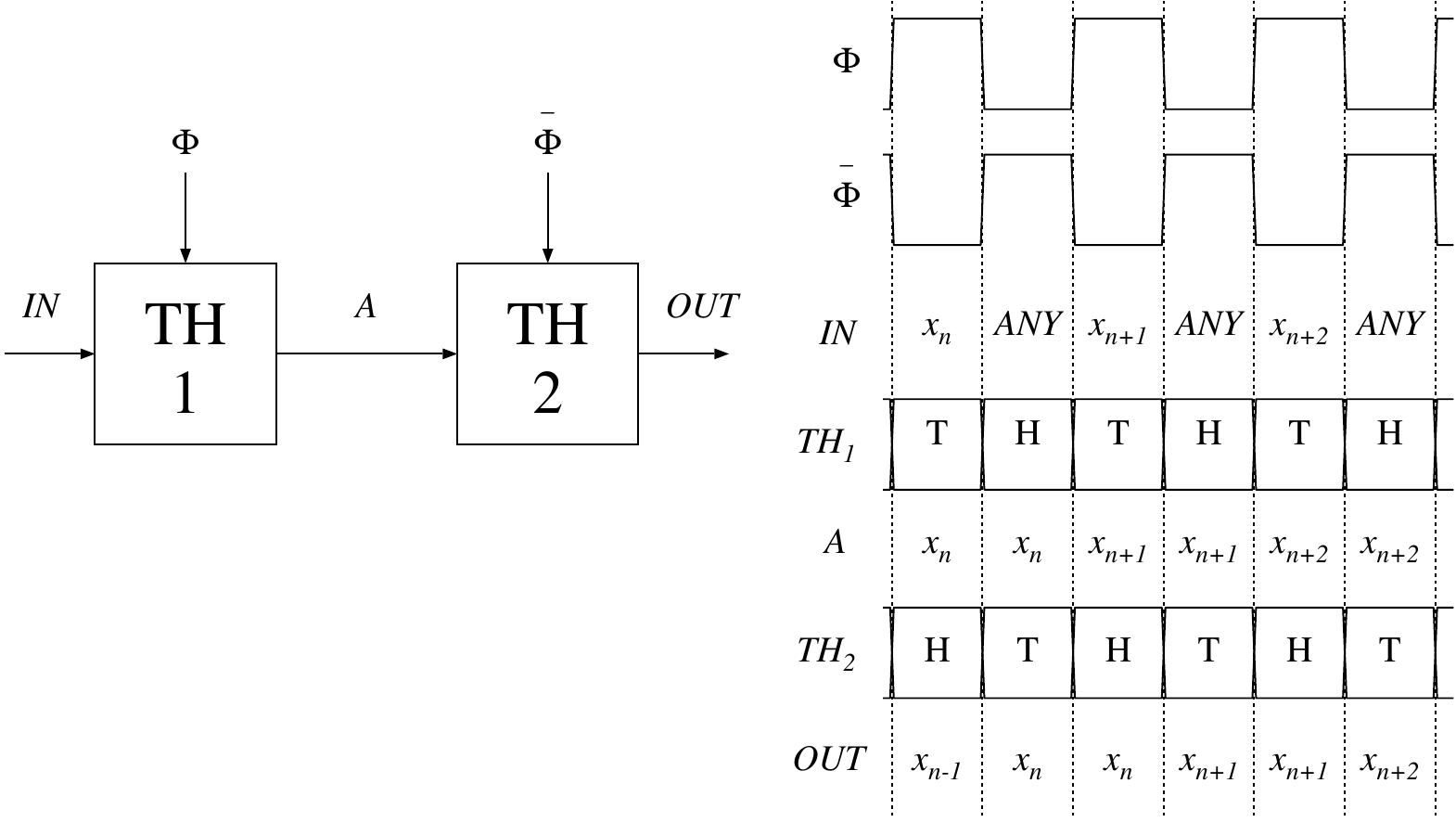}
  \caption{\label{fig:register1}%
    Timings of an analog register designed by cascading two TH
    elements (``T'' = track phase, ``H'' = hold phase).}
\end{figure}

\begin{figure}[ht]
  \centering\includegraphics[width=0.95\lw]{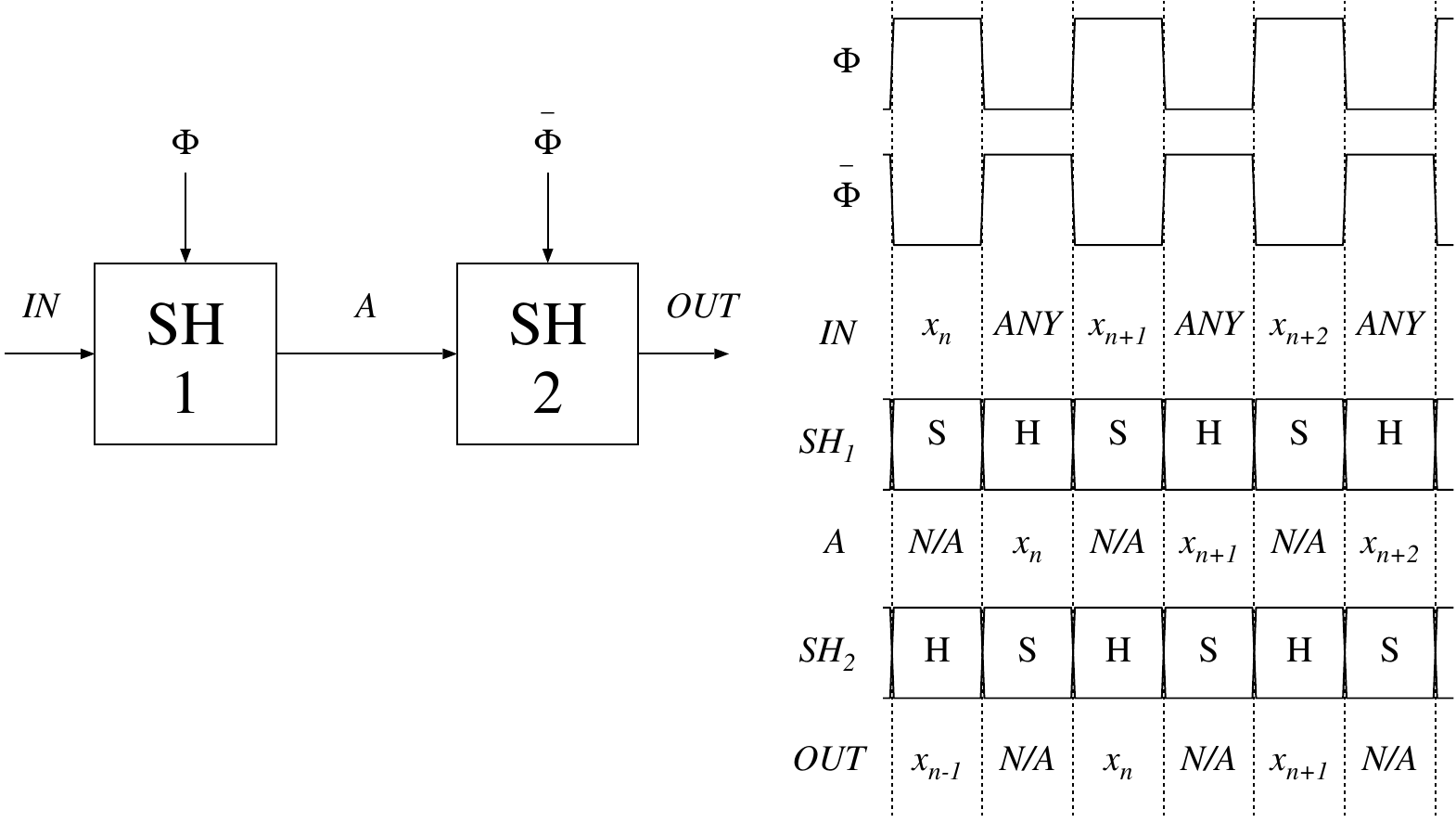}
  \caption{\label{fig:register2}%
    Timings of an analog register designed by cascading two SH
    elements (keys as in the previous figure, plus ``S'' = sample
    phase, ``N/A'' = not available).}
\end{figure}

\begin{figure}[ht]
  \centering\includegraphics[width=0.95\lw]{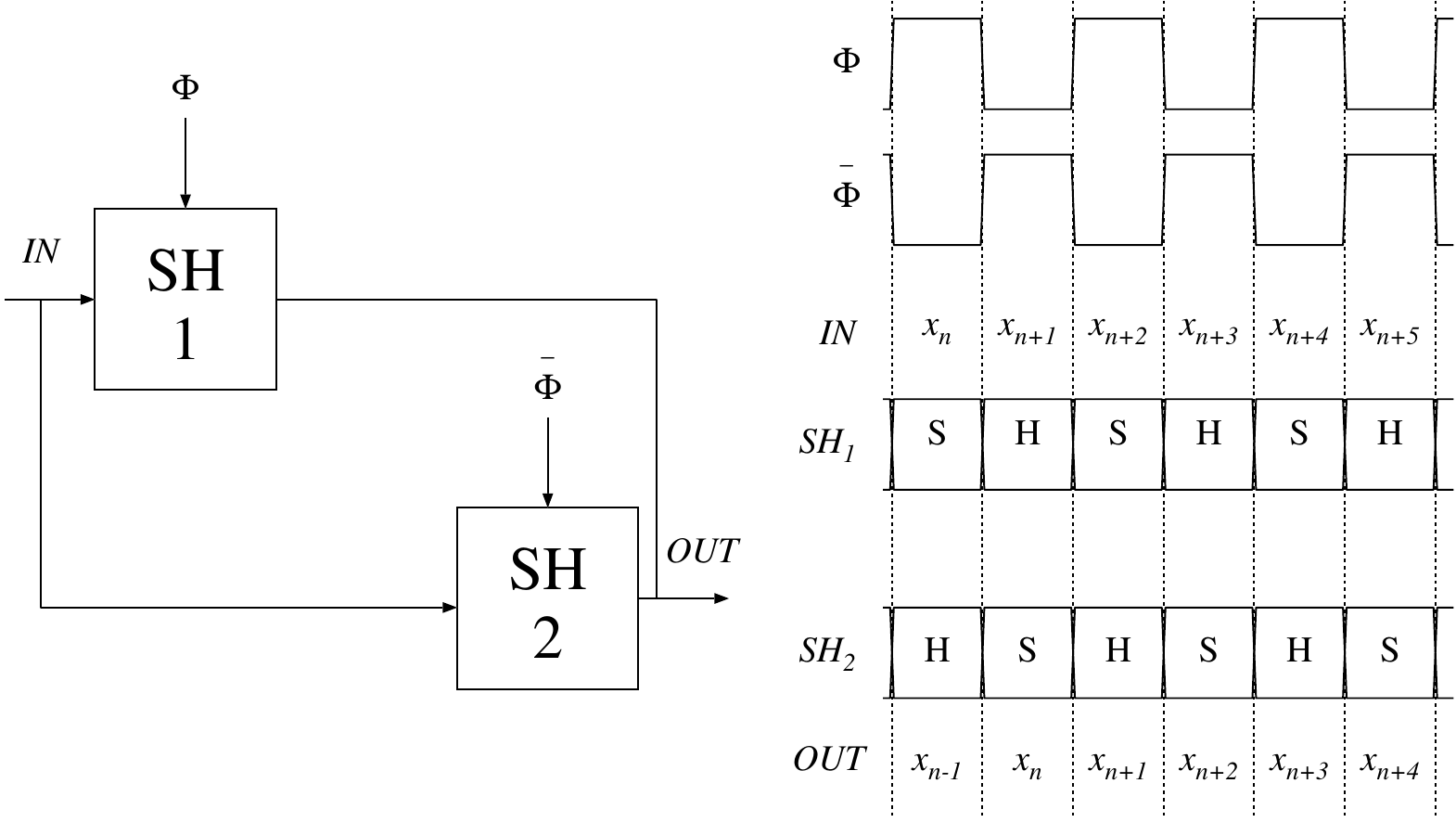}
  \caption{\label{fig:register3}%
    Timings of an analog register designed by operating two SH
    elements in anti-parallel (keys as in the previous figures).}
\end{figure}

For what concerns the register architecture and the preference to
SH\emph{s} or TH\emph{s}, it can be noticed that correct operation can
be obtained either by cascading TH elements (as shown in
Figures~\ref{fig:TTMchaos} and~\ref{fig:register1}),
by cascading SH elements (Figure~\ref{fig:register2}),
or by having two SH elements operating in anti-parallel
(Figure~\ref{fig:register3}).
Note that in the current mode design approach, SH circuits
(\emph{dynamic current mirrors}) show many advantages (e.g.\@ the
absence of device matching issues) over TH\emph{s} and are generally
preferred.  The figures show also the timing diagrams where the
differences are justified by the fact that TH circuits can provide a
valid output during the sampling phase, while SH\emph{s} can not.
From the over-mentioned figures it is evident that the cascaded
circuits can process a new sample at every clock period, while the
anti-parallel one is capable of a new sample every \emph{half} clock
period and is inherently faster. Of course, in a practical circuit
this is important only if it is the analog register itself to be the
speed limiting factor, yet in chaotic circuits this is often the case.

Thus, was it not for other design issues, there would be no doubt for
preferring the anti-parallel SH solution. Actually, this is where the
third degree of freedom, i.e.\@ the choice of a particular SH circuit
comes into play.

As mentioned above, in-loop errors must be kept extremely low, so that
optimized SH architectures must be selected. When discussing the
accuracy of analog memory element a distinction among \emph{signal
  independent} and \emph{signal dependent} errors comes natural, so
that, in current mode operation, one writes:
\begin{equation}
  I\DMs{out_n}=I\DMs{in_{n-1}} + f(I\DMs{in_{n-1}}) + \sigma
\end{equation}
where the three addends on the righthandside represent the ideal
behaviour, the signal dependent and the signal independent error
respectively. Discrete time chaotic circuit are known to be more
sensitive to signal dependent errors \cite{Callegari:NOLTA98}.

\begin{figure}[ht]
 \centering\includegraphics[width=0.8\lw]{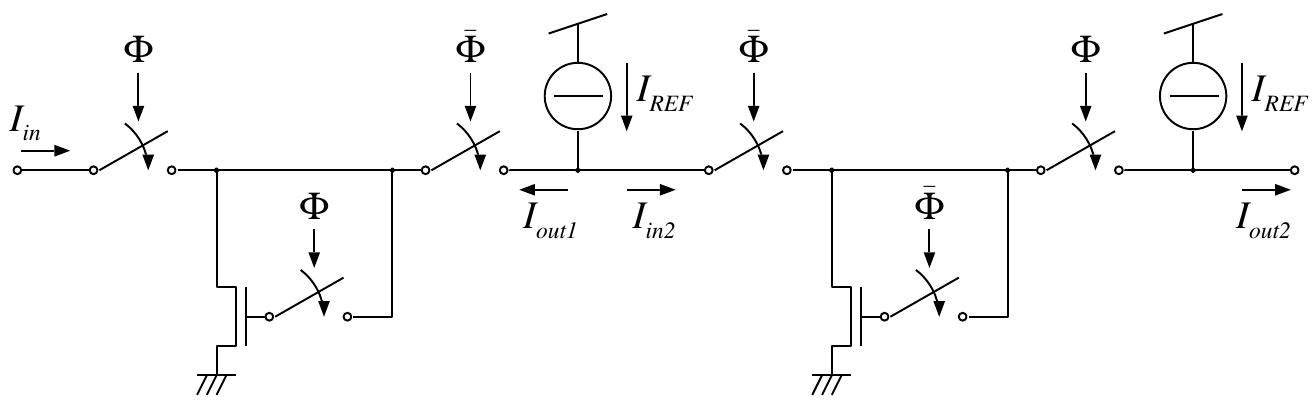}
  \caption{\label{fig:samplecascade}%
    A sample analog register which cancels signal independent errors
    exploiting SH cascading.}
\end{figure}

In current mode operation, the cascade of two SH (or TH) stages has
the property of allowing the cancellation of signal independent
errors, as long as the matching among the two stages is sufficiently
good. For instance, one can easily arrange things in the spirit of the
sample circuit of Figure~\ref{fig:samplecascade}, where:
\begin{equation}
  \begin{cases}
    I\DMs{out1} =& I\DMs{in} + f(I\DMs{in}) + \sigma\\
    I\DMs{in2} =& I\DMs{REF} - I\DMs{out1}\\
    I\DMs{out2} =& I\DMs{in2} + f(I\DMs{in2}) + \sigma\\
    I\DMs{out} =& I\DMs{REF} - I\DMs{out2}    
  \end{cases}
\end{equation}
Note $I\DMs{REF}$ is a suitably large current, and that the
time-indexes $n$, $n-1$ etc. have been omitted for simplicity. The
output current is thus
\begin{equation}
I\DMs{out}=I\DMs{in}+g(I\DMs{in})
\end{equation}
where $g(I\DMs{in})$ is the overall signal dependent error.

Hence there is an opportunity of selecting SH circuits specially
optimized for reducing signal dependent errors, then relying on SH
cascading for the reduction of signal independent ones. In this way
the signal dependent errors to which chaotic circuits are most
sensitive can be addressed at their best, within each SH unit, while
less critic signal independent errors can be dealt with at the
register level exploiting the SH matching properties.  For instance,
in \cite{Callegari:ECCTD99} it is proposed the adoption of S\Us{2}I
SH\emph{s} \cite{Hughes:ISCAS93} which have the property of dealing
extremely well with signal dependent errors regardless of matching,
while leaving a relatively large residual signal independent error.

Because of the signal independent error cancellation property, the
cascaded SH analog register architecture is often preferable to the
anti-parallel one. Unfortunately, this does normally mean sacrificing
the operating speed of the anti-parallel topology. In the following,
an interleaving technique is proposed which allows retaining the
advantages of both worlds.

\section*{Interleaving}
\begin{figure}[ht]
  \centering\includegraphics[width=0.95\lw]{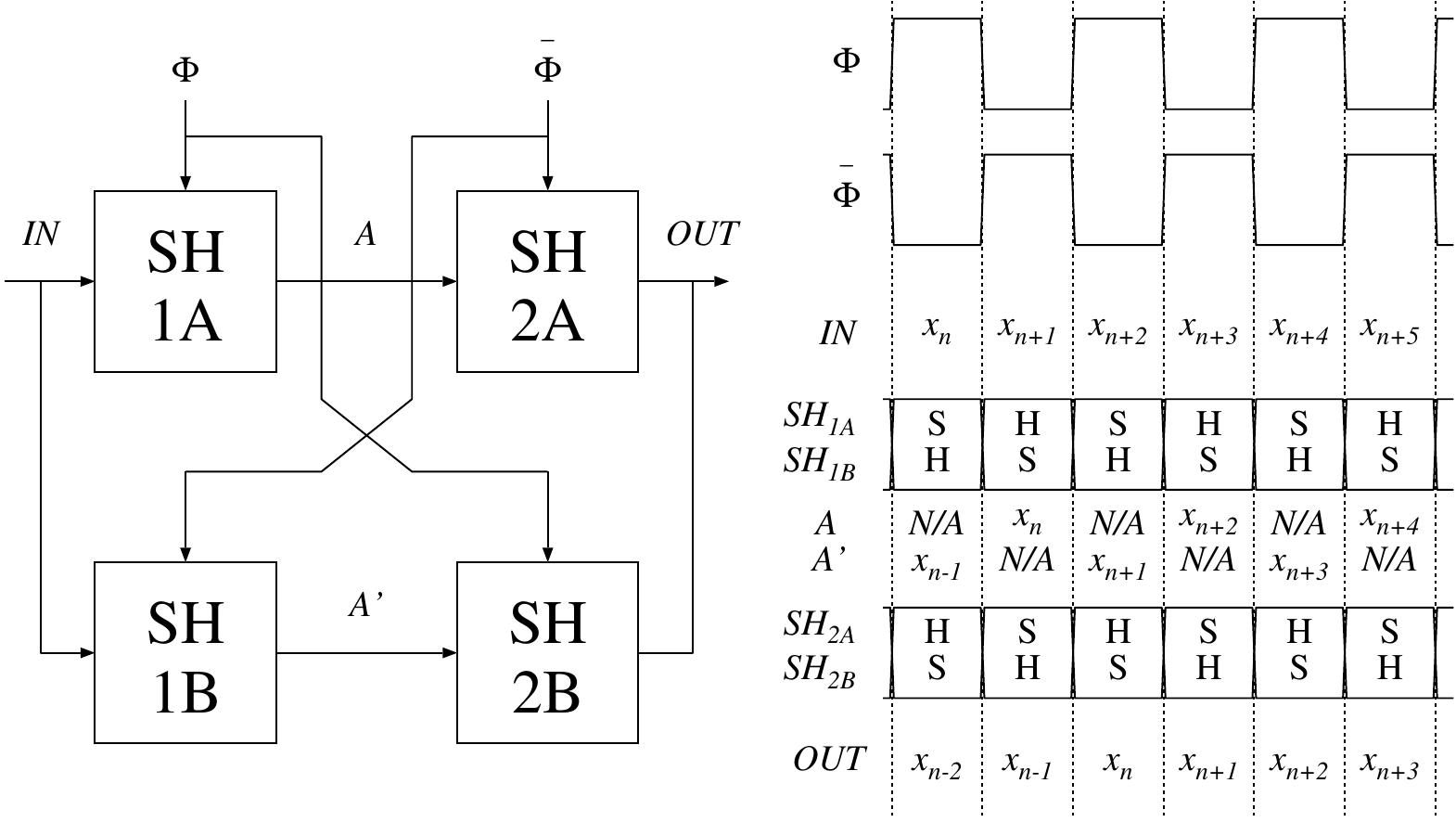}
  \caption{\label{fig:register4}%
    Timings of an analog register designed by operating two whole
    cascaded-SH analog registers in anti-parallel (keys as in the
    previous figures).}
\end{figure}

If two cascaded-SH analog registers are \emph{themselves} connected in
anti-parallel, then the topology and the timing diagrams shown in
Figure~\ref{fig:register4} are obtained.

Note that this architecture provides cancellation of SH signal
independent errors. Furthermore, it has the same throughput as the
architecture in Figure~\ref{fig:register3}, i.e.\@ a new sample every
half clock cycle. The only difference is the associated delay: in the
topology of Figure~\ref{fig:register3} the output is delayed half a
clock cycle with regard to input, while in the topology of
Figure~\ref{fig:register4} the delay is a whole clock cycle.

This means that if the architecture shown in
Figure~\ref{fig:register4} is substituted for the analog memory in the
discrete time system given in Figure~\ref{fig:TTMchaos}, still one
\emph{cannot} obtain a chaotic circuit providing a new chaotic sample
every half clock cycle. However, \emph{two independent identical}
chaotic sources showing interleaved operation can be obtained, as it
will be illustrated shortly. 

Let us accept that these two systems exist, and name them `(A)' and
`(B)'. Suppose that the sample $x_0$ in Figure~\ref{fig:register4}
corresponds to the initial condition $x\Us{(A)}_0$ of system `(A)' and
that $x_1$ corresponds to the initial condition $x\Us{(B)}_0$ of
system `(B)'. If the clock period is $T$ and at $t=0$ the analog
register outputs $x_0$, then from $t=0$ to $t=T/2$ the map circuit
operates on $x\Us{(A)}_0=x_0$ producing $x\Us{(A)}_1=f(x\Us{(A)}_0)$.
At the same time, the analog register samples $x_2$, hence
$x_2=x\Us{(A)}_1$. From $t=T/2$ to $t=T$ the analog register provides
$x_1$, causing the map circuit to operate on $x\Us{(B)}_0=x_1$. Thus
the map provides $x\Us{(B)}_1$ and the register samples
$x_3=x\Us{(B)}_1$. From $t=T$ to $t=3/2 T$, the register outputs
$x_2$, i.e.\@ $x\Us{(A)}_1$, the map computes $x\Us{(A)}_2$, which is
in turn sampled as $x_4$. From $t=3/2 T$ to $t=2T$, the register
outputs $x_3$, i.e.\@ $x\Us{(B)}_1$ and the map computes $x\Us{(B)}_2$
which is sampled as $x_5$ and so on.

It is self-evident how two different chaotic trajectories emerging
from the same map are kept separated in a time-division fashion. In
general terms, the information stored on the \emph{interleaved} analog
register can be related to the state of the chaotic systems as:
\begin{equation}
  x_n=
  \begin{cases}
    x\Us{(A)}_{n/2} &\quad\text{if $n$ is even}\\
    x\Us{(B)}_{(n-1)/2} &\quad\text{if $n$ is odd}
  \end{cases}
\end{equation}
In other words, the SH\emph{s} marked `A' in
Figure~\ref{fig:register4} store and transfer the state of system `A',
while those marked `B' store the state of system `B'.

The two systems are obviously independent, since there is no exchange
of information among the two. What remains to be seen is if the
sequences $x\Us{(A)}_n$ and $x\Us{(B)}_n$ are themselves independent
in a \emph{cross-correlation} sense. What if at startup the initial
conditions $x_0$ and $x_1$ are identical? Being the dynamic chaotic,
there is actually no worry about synchronization. In fact,
\emph{sensitivity to initial conditions} assures that the trajectories
of system `(A)' and `(B)' cannot help rapidly diverging because of the
unavoidable mismatch in the initial conditions and the effects of
noise. Note that this would be true of any two independent chaotic
sources based on the same map, yet this arrangement is particularly
convenient, as it allows to realize two identical chaotic sources
while sharing a large part of their hardware (the whole map circuit)
and to obtain an output sample every half clock cycle.

At this point, it is interesting to consider the statistical
properties of the interleaved sequence at the output of the analog
register, and to verify whether they can be useful to some
applications.

If we consider the sequence $x_n$, its probability density function
(PDF) is of course the same as that of $x\Us{(A)}_n$ (or
$x\Us{(B)}_n$). Hence, interleaving does not influence the first order
statistics. On the contrary, the autocorrelation is obviously
affected. In fact we have:
\begin{equation}
  \Phi_{x,x}(n)=
  \begin{cases}
    \Phi_{x\Us{(A)},x\Us{(A)}}(n/2) &\quad\text{if $n$ is even}\\
    \mu^2 &\quad\text{if $n$ is odd}
  \end{cases}
\end{equation}
where $\mu$ is the average (or DC component) of the sequences
$x\Us{(A)}_n$ and $x\Us{(B)}_n$, that we shall assume to be zero for
simplicity.  If we take the Fourier transform of the autocorrelation
we obtain the power density spectrum, resulting in:
\begin{equation}
  X(j\omega)=\sum_{n=-\infty}^{\infty} \Phi_{x,x}(n)
  e^{-j\omega\frac{T}{2}n}
\end{equation}
which can be rewritten as
\begin{equation}
  \begin{split}
    X(j\omega) &=
    \sum_{\substack{n=-\infty\\\text{$n$ even}}}^{\infty} \Phi_{x,x}(n)
    e^{-j\omega\frac{T}{2}n} +\\
    &\quad+
    \sum_{\substack{n=-\infty\\\text{$n$ odd}}}^{\infty} \Phi_{x,x}(n)
    e^{-j\omega\frac{T}{2}n} =\\
    &=
    \sum_{m=-\infty}^{\infty} \Phi_{x,x}(2m)
    e^{-j\omega\frac{T}{2}2m}\\
    &=
    \sum_{m=-\infty}^{\infty} \Phi_{x\Us{(A)},x\Us{(A)}}(m)
    e^{-j\omega T m} = X\Us{(A)}(j\omega)
 \end{split}
\end{equation}
where $T$ is the clock period. Hence, being a periodic function, the
power density spectrum is not affected.  Higher order moments, which
are harder but not impossible to compute, are generally all affected,
but they have a lower impact on typical applications.

\section*{Application example}
\begin{figure}[ht]
  \centering\includegraphics[width=0.97\lw]{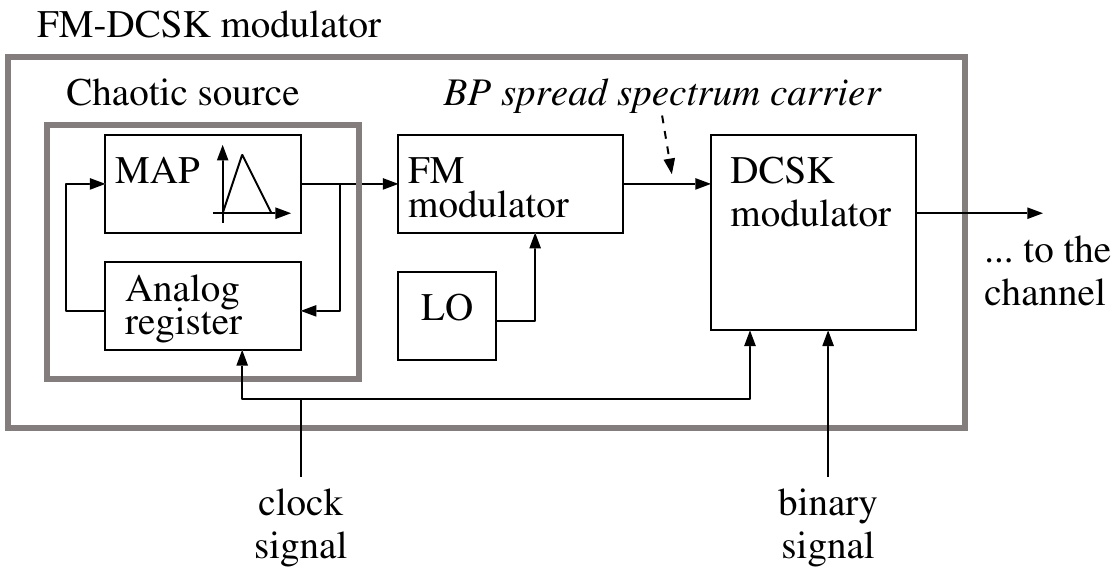}
  \caption{\label{fig:fmdcsk}%
    FM-DCSK modulator and transmission unit.}
\end{figure}

In order to show an example in which an interleaved chaotic source is
adopted instead of a traditional one, we shall consider an FM-DCSK
communication system \cite{Kolumban:TIEICE-81A-9,Kolumban:ECCTD99}. In
an FM-DCSK modulator, a discrete time uniform PDF random/chaotic
source is used as the input to an FM modulator. In this way a
spread-spectrum band-pass carrier is obtained to be fed to a further
DCSK modulator, as shown in Figure~\ref{fig:fmdcsk}. Ideally the
spread spectrum carrier should be characterized by a bandwidth limited
to an interval and by a uniform power density spectrum over that
interval.

\begin{figure}[ht]
  \centering
  \includegraphics[width=0.49\lw]{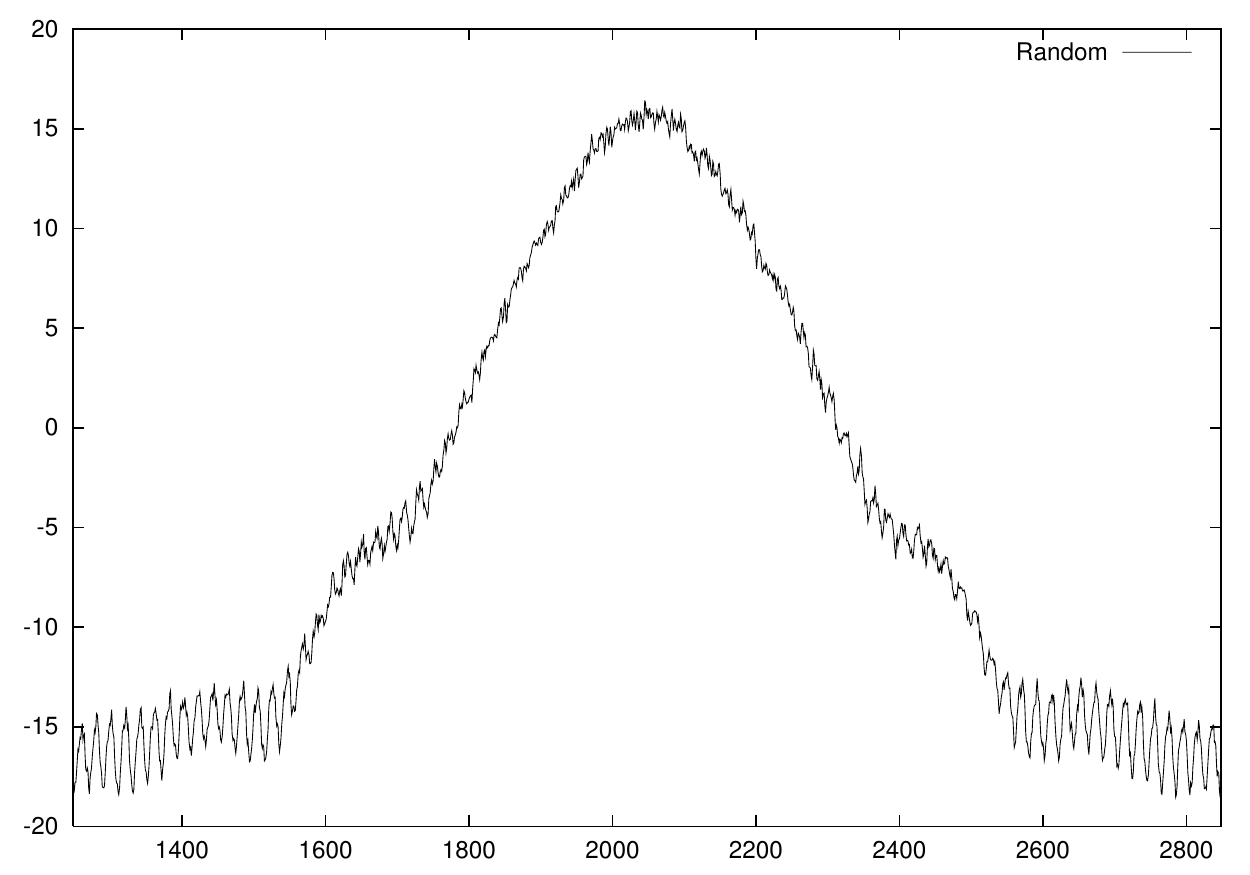}
  \includegraphics[width=0.49\lw]{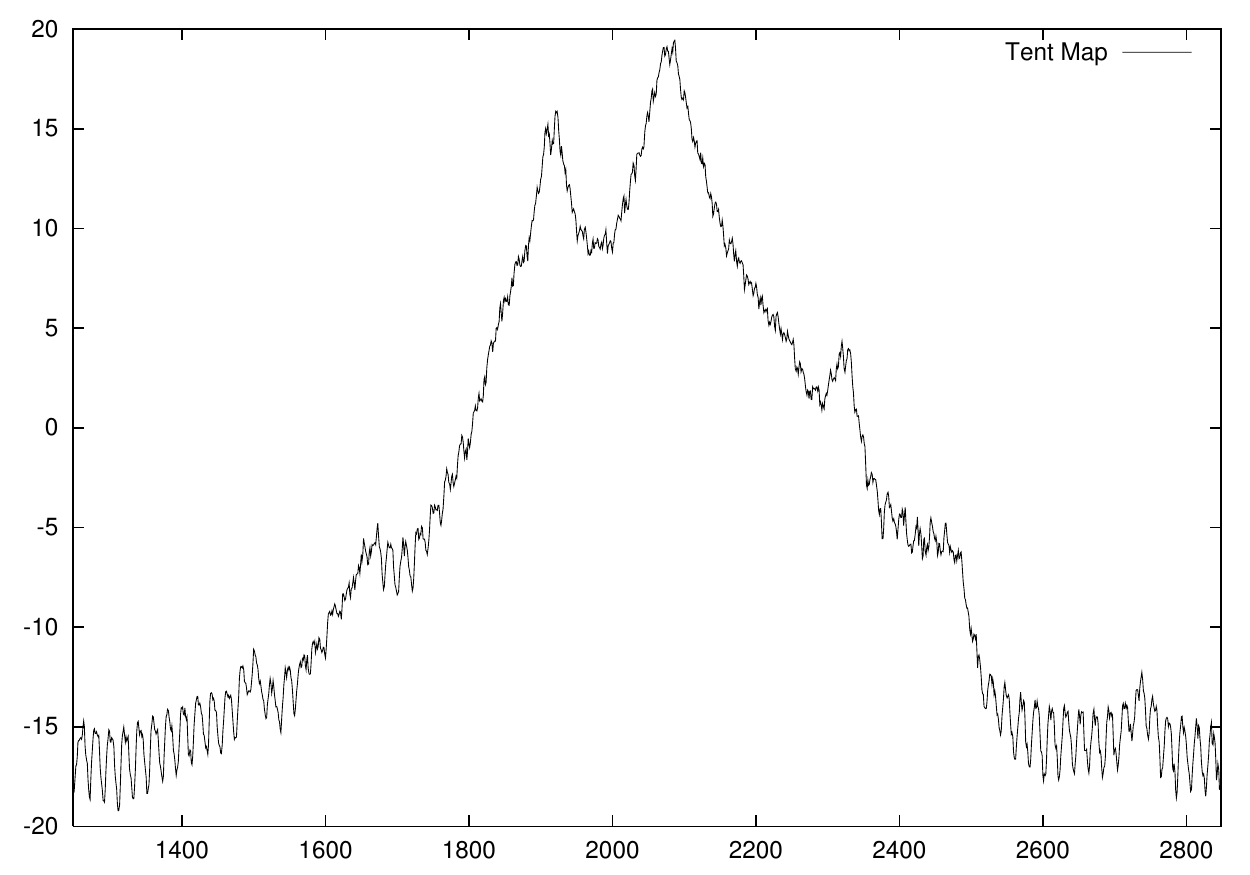}\\
  \includegraphics[width=0.49\lw]{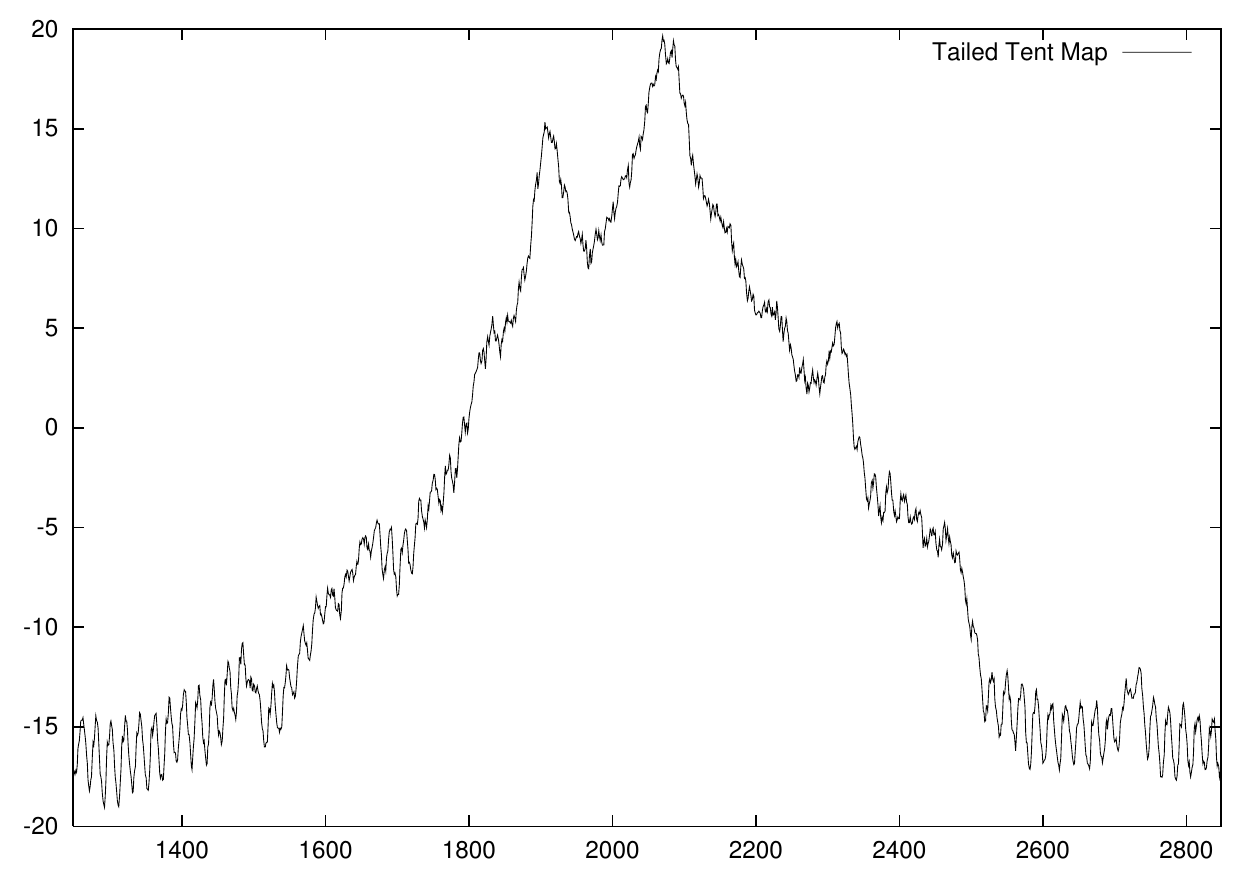}
  \includegraphics[width=0.49\lw]{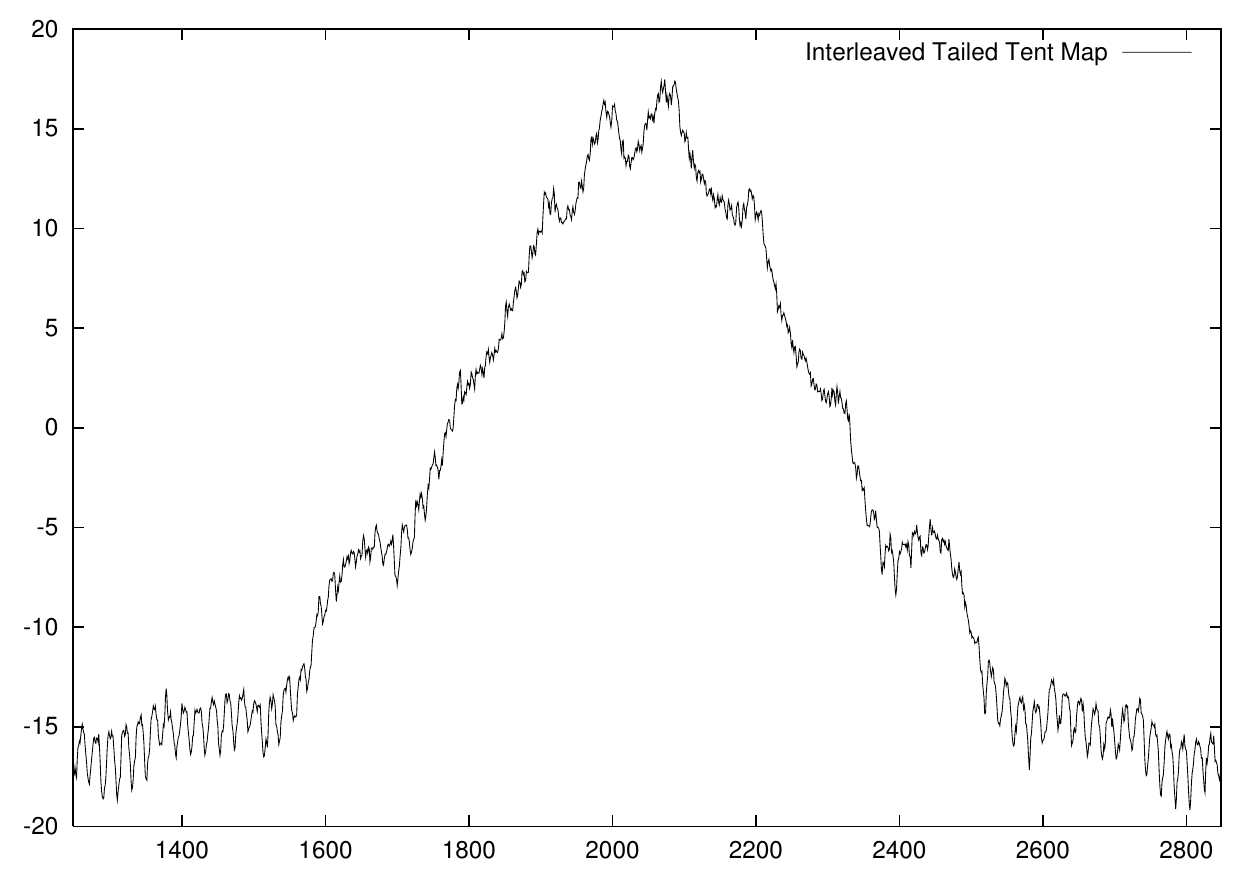}
  \caption{\label{fig:spectrum}%
    Spectrum of the FM-DCSK signal for \emph{fast} FM-DCSK system. Top
    left corresponds to random data fed to the FM modulator; top right
    to a chaotic stream produced by a tent map based system; bottom
    left to a chaotic stream produced by a tailed-tent-map system and
    bottom right by an interleaved tailed tent map system. Note that
    on the frequency axis value 2048 corresponds to
    \Unit{100}{\mega\hertz} and that the amplitude axis is in
    \decibel.}
\end{figure}

If one looks at a practical case of the so called \emph{fast FM-DCSK},
where a random source is used to drive the FM modulator, a
spectrum as the one in Figure~\ref{fig:spectrum} (top-left) is obtained for
the signal out of the DCSK modulator. The non-uniform power
distribution is clearly given by the FM modulation index which is used
in practical \emph{fast} FM-DCSK systems.

If a uniform PDF white-spectrum chaotic source is used instead of a
random source, then the spectral characteristics of the DCSK carrier
are somehow slightly deteriorated. Typically, the closer to one the
mixing rate $r\DMs{mix}$ of the chaotic process
\cite{Lasota:CFN-1995}, the more perceptible the influence on the
spectrum. Easy to implement chaotic systems (such as tent-map based
ones) show mixing rates which allow to perceive the spectral
deterioration, but still allow a satisfying behaviour. Top-right plot
in Figure~\ref{fig:spectrum} shows the spectral properties of the
FM-DCSK signal when a chaotic tent-map base process is used to feed
the FM modulator.

In practice, it is common to use chaotic models where particular
strategies are adopted to enhance the implementation robustness. An
example is offered by tailed tent map (TTM) systems
\cite{Callegari:ISCAS97}, such as:
\begin{equation}
  x_{n+1}=1-2|x-(1-\theta)/2|+\max(x-1+\theta,0)
\end{equation}
where $\theta$ is a control parameter in $(0,1/2)$ usually taken very
close to zero. By adopting a TTM chaos generator ($\theta=0.05$) the
FM-DCSK signal spectrum is modified as shown by the bottom left plot
in Figure~\ref{fig:spectrum}.

Finally, the bottom right plot shows the spectral properties of a DCSK
signal generated using an interleaved TTM based chaotic source.  In
this case not only the applicability of the interleaving technique is
verified: a slightly better behaviour is also obtained. This
improvement is due to the effects of the interleaving technique on the
higher order moments of the chaotic process. However note that the
impact on the overall behaviour of a complete FM-DCSK system is almost
negligible.  However, the possibility of using an interleaved source
is extremely interesting, since it allows to design chaotic sources
capable of reaching the data rate requirements of FM-DCSK modulators.
As a reference value, consider that typical systems designed so far
require \Unit{20}{\mega\sample\per\second}.
 
\section*{Sample circuit}
In order to show the effectiveness of the proposed technique, a sample
circuit has been designed and simulated. For this testing a well
established \Unit{0.8}{\micro\metre} CMOS technology has been adopted.
The circuit has been designed to reach an output data rate sufficient
for the FM-DCSK application illustrated above.

The schematic that we illustrate is based on
\cite{Callegari:ECCTD99}, where S\Us{2}I dynamic mirrors
\cite{Hughes:ISCAS93} are used for enhanced speed and accuracy. The
TTM is chosen as the system nonlinear function. 

\begin{figure}[ht]
  \centering
  \resizebox{0.95\lw}{!}{\includegraphics{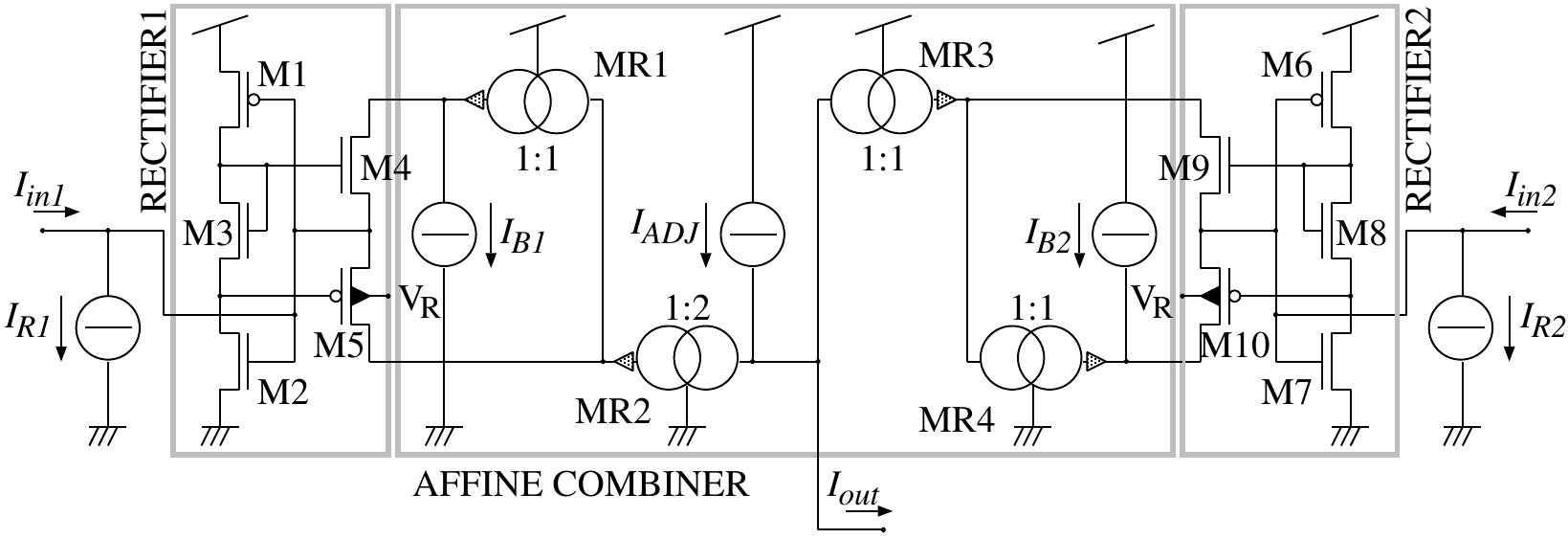}}
  \caption{\label{fig:ttm-c}%
    Current mode TTM circuit.}
\end{figure}

The map circuit is shown in Figure~\ref{fig:ttm-c}.
Note that it requires two identical input currents to perform
concurrently the comparisons necessary to evaluate the relative
position of the input and the breakpoints. For this evaluation active
circuits are used, so that the main feature of the circuit is that the
voltage at the inputs can be kept almost constant.  In
\cite{Callegari:ECCTD99}, a convenient way is suggested
to make this voltage $V_R$ generally available by means of a dummy rectifier.
This allows to use $V_R$ in other parts of the circuit, for instance
to connect the wells of M5 and M10, cancelling the body effect and
improving performance.

Reference currents $I_{R1}$ and $I_{R2}$ are set respectively to
$-2.5$ and \Unit{45}{\micro\ampere} and set the breakpoint positions
($\theta=1/20$).  Currents $I_{B1}$ and $I_{B2}$
(\Unit{2.5}{\micro\ampere}) speed up signal processing by never
allowing mirrors to operate on null currents. Finally,
$I_{\text{ADJ}}$ (\Unit{57.5}{\micro\ampere}) fixes the output offset
and sets the system invariant set to \Unit{[-50,+50]}{\micro\ampere}.
Current mirrors are all \emph{high swing cascode} and gate areas of
transistors that could introduce \emph{matching errors} are always
non-minimal (typically > \Unit{40}{\mu\metre\squared}).

\begin{figure*}[ht]
  \centering
  \resizebox{0.75\lw}{!}{\includegraphics{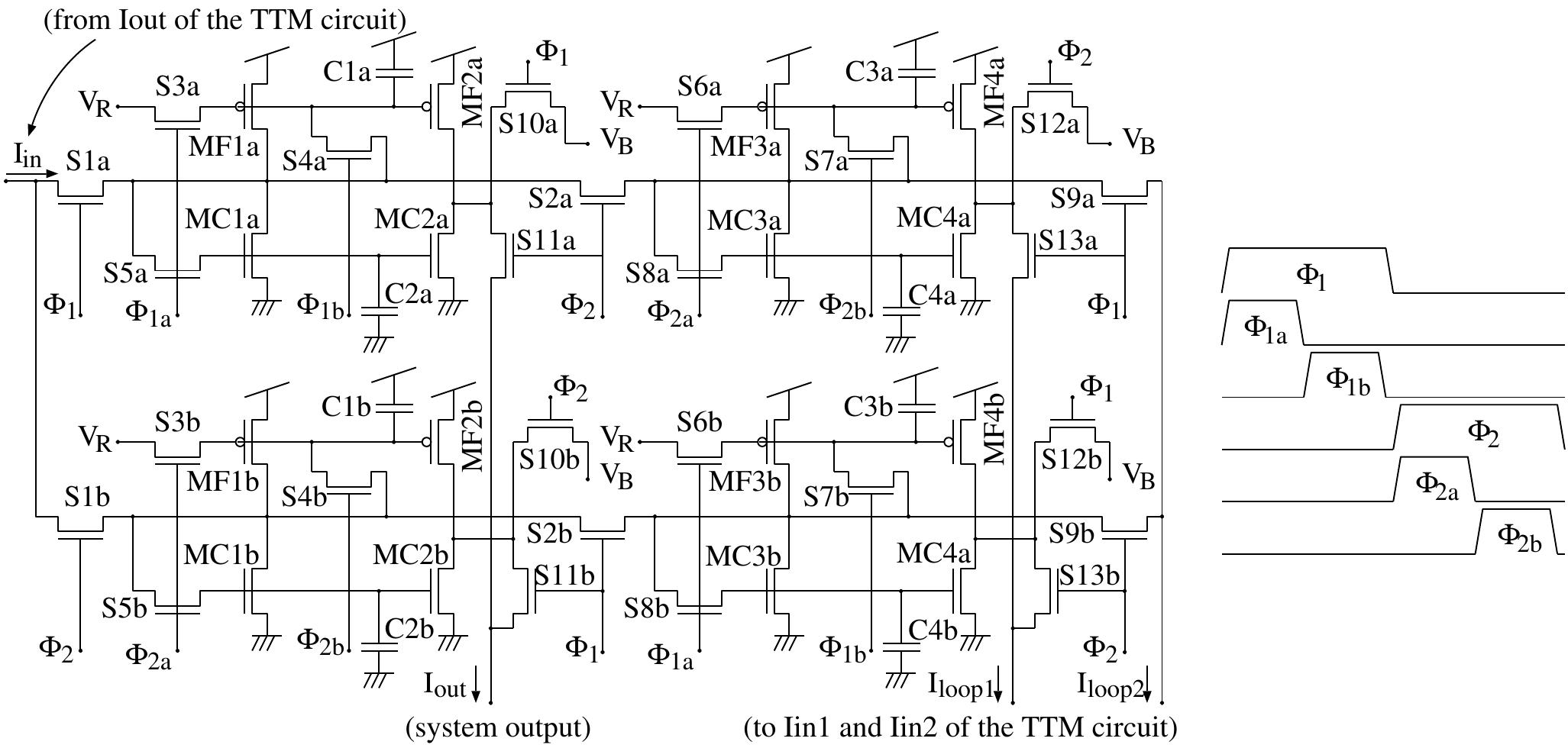}}
  \caption{\label{fig:muxs2i}%
    Interleaved analog delay elements and timing.}
\end{figure*}

The analog delay is shown in Figure~\ref{fig:muxs2i}.
This already comprises two interleaved memory units, each of them
built up of two S\Us2I dynamic current mirrors \cite{Hughes:ISCAS93}.
The clocking scheme is as for a typical S\Us2I system and the unit
embeds mirroring to provide currents to both the map circuit
inputs and the chaotic system output.  For convenience in
Figure~\ref{fig:muxs2i} the two interleaved analog memories are
distinguished by the postfixes ``a'' or ``b'' applied to all the
relevant devices, while the many switches are named S\emph{xxx}.

Note that (as pointed out in \cite{Callegari:ECCTD99}) for correct interfacing
to the map circuit, the voltage $V_R$ mentioned above must be used as the
reference voltage for the S\Us2I dynamic mirrors in order to exploit the low
input impedance of the map unit (in the figure, $V_B$ is simply a buffered
version of $V_R$).  This circuit ideally propagates the state variables without
introducing signal dependent errors, which is an important condition for the
accurate operation of the chaotic system. Only the output variable is subject
to a non-negligible sampling error, which is an offset and thus easy to deal
with.

\begin{figure}[ht]
  \centering
  \resizebox{0.95\linewidth}{!}{\includegraphics{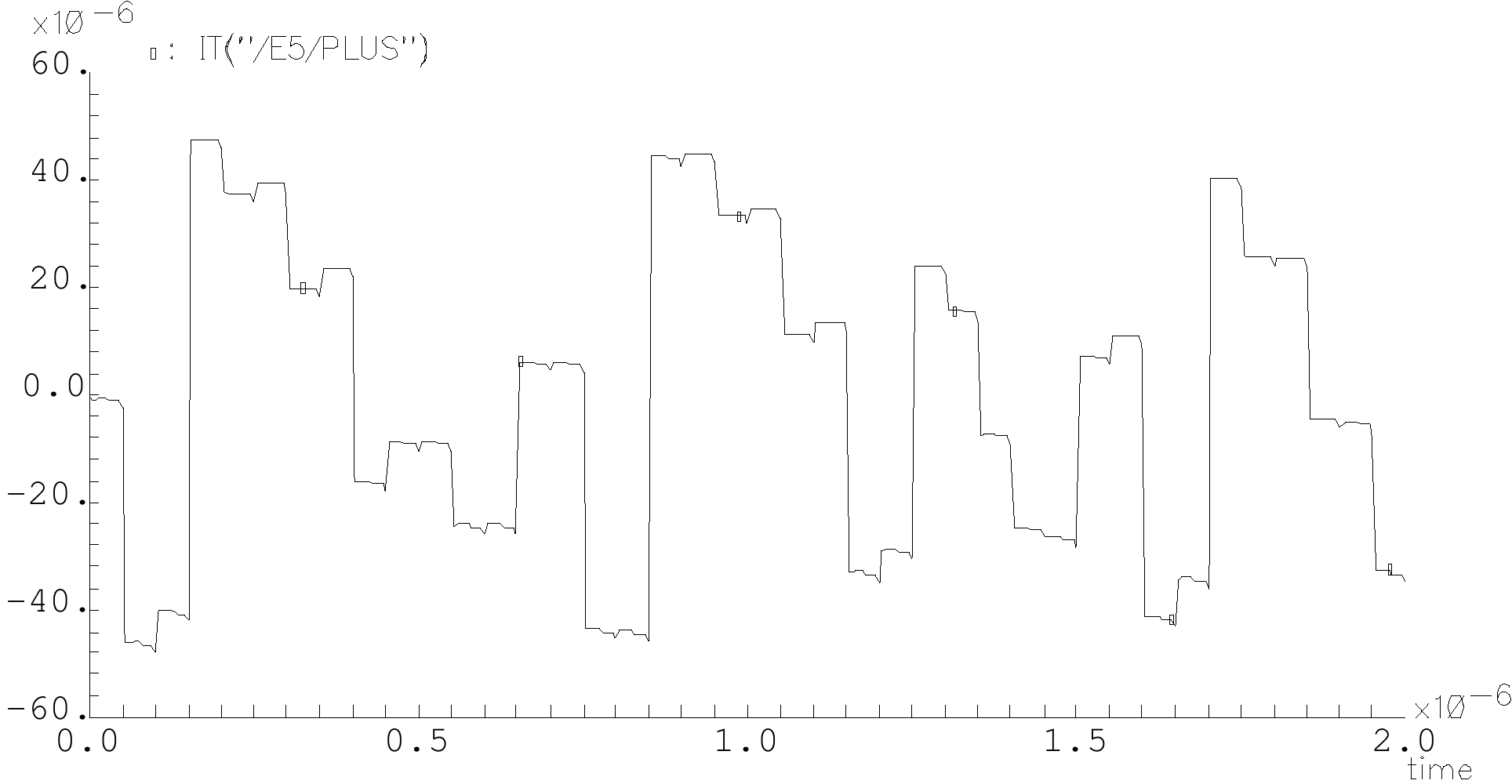}}
  \caption{\label{fig:operation}%
    \itshape The system output current as simulated by SPICE.}
\end{figure}

Figure~\ref{fig:operation}
shows the evolution of the output variable when the system is
simulated using SPICE with accurate analog device models (BSIM 3).
The output data rate is \Unit{20}{\mega\sample\per\second}.

\begin{figure}[ht]
  \centering\includegraphics[width=0.8\lw]{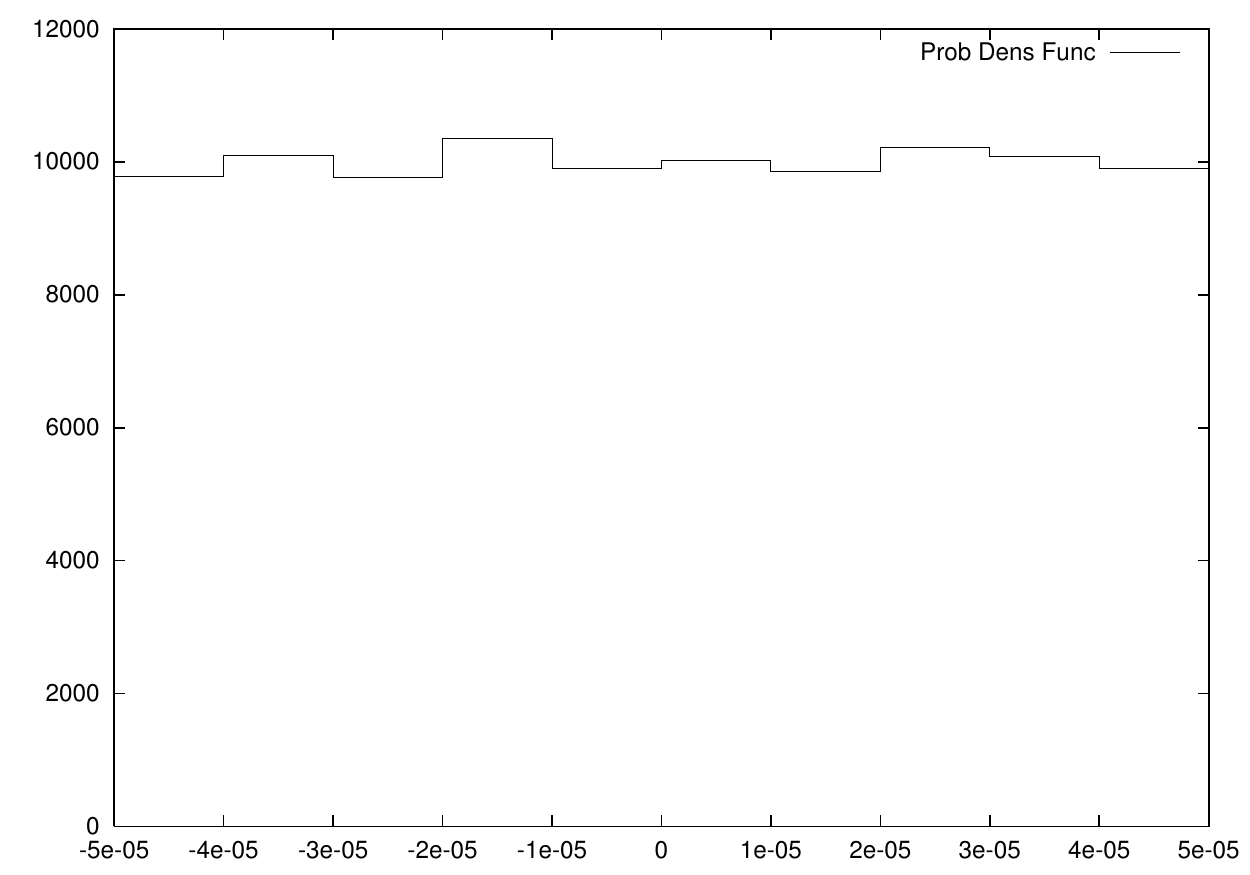}
  \caption{\label{fig:intlvpdf}%
    Estimation of the analog register variable $x_n$ PDF for
    the presented circuit.}
\end{figure}

\begin{figure}[ht]
  \centering\includegraphics[width=0.8\lw]{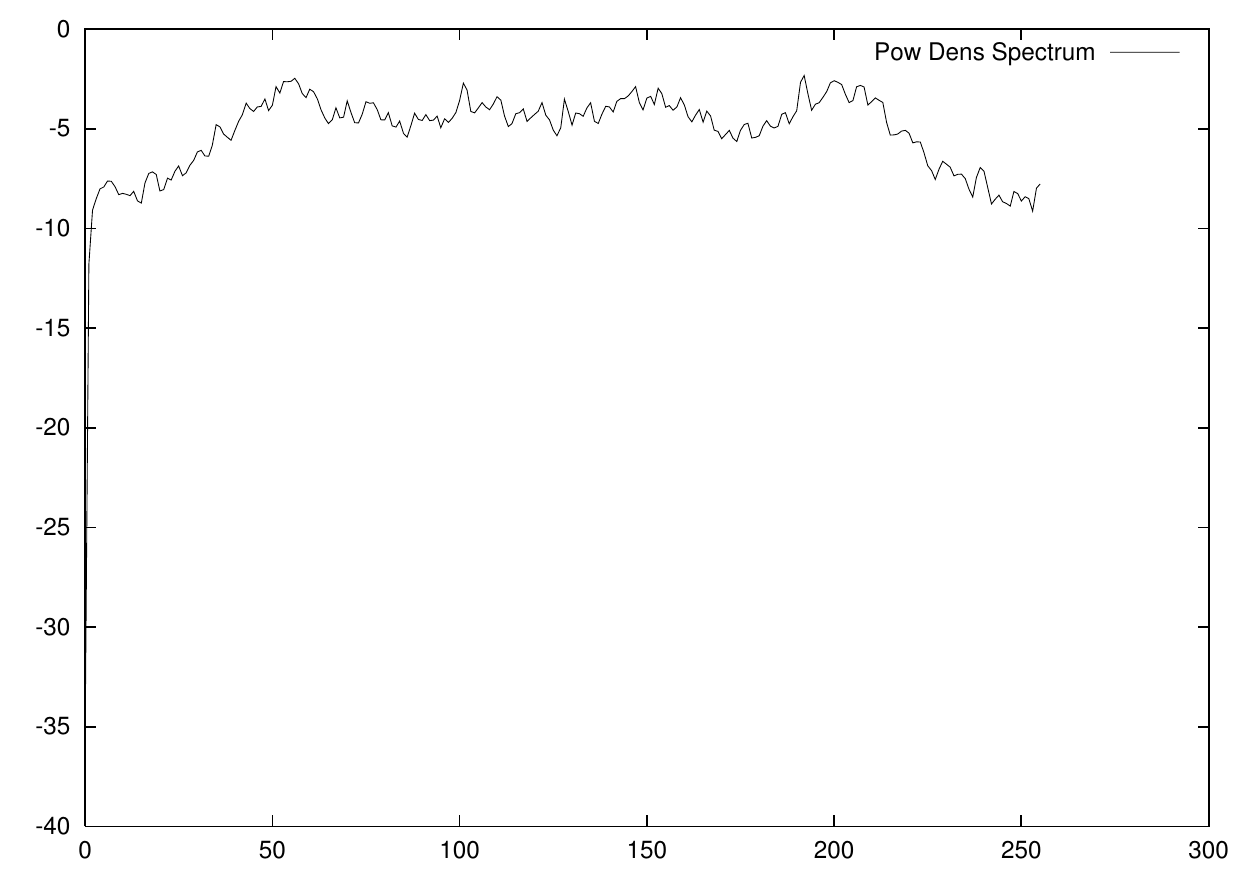}
  \caption{\label{fig:intlvspec}%
    Estimation of the power density spectrum relative analog
    register variable $x_n$ for the presented circuit. On the $x$-axis
    the value $256$ corresponds to the Shannon frequency
    (\Unit{10}{\mega\hertz}). The $y$-axis is in \decibel.}
\end{figure}

The observation of longer sequences has shown that the unavoidable
coupling among the two interleaved chaotic systems due to circuit
parasitic does not produce perceivable effects such as synchronization.
Direct estimation of statistical properties from spice data has proven
hard, due to the inefficiency of SPICE time-step algorithms when
dealing with S\Us2I circuits. In practice, it has not been possible to
obtain long enough sequences for reliable statistics
estimation. Nonetheless, a mix of Spice and behavioural simulations
has allowed to obtain the plots in Figures~\ref{fig:intlvpdf}
and~~\ref{fig:intlvspec}. The first refers to the probability density
function and the second to the power density spectrum of the
state variable which is propagated through the analog register.

\section*{Conclusions}
In this paper, the use of an interleaving technique applied to
discrete time chaotic sources has been proposed. Its feasibility has
been analytically shown and the usability of the so obtained
chaotic sources for a sample application (FM-DCSK communication) has
been verified by simulations at the model level.

To validate the possibility of exploiting the proposed technique at
the hardware level, a CMOS circuit has been designed using a
conventional \Unit{0.8}{\micro\metre} technology. As far as
simulations have been ran, the circuit has shown operation in
accordance with the expected behaviour. Some data is still lacking
about the estimation of the spectral properties of the hardware
interleaved chaotic generator.

Currently work is in progress, both in order to obtain good estimation
of the statistical properties of the system directly from circuit
simulation and to evaluate the usability of interleaved chaotic
sources to other application fields.

\section*{The authors}
Sergio Callegari and Riccardo Rovatti are with the Department of
Electronics, Computer Science and Systems (DEIS) of the University of
Bologna, Viale Risorgimento 2, 40137 - Bologna (ITALY).\\
\parbox{2cm}{E-mail:}\mailto{scallegari@deis.unibo.it}\\
\hspace*{2cm}\mailto{rrovatti@deis.unibo.it}

Gianluca Setti is with the Department of Information Science (DI) of
the University of Ferrara, Via Saragat 1, 44100 - Ferrara (ITALY).\\
\parbox{2cm}{E-mail:}\mailto{gsetti@ing.unife.it}

\bibliography{macros,analog,chaos_theory,chaos_circuits,chaos_applications,%
  pram}

\begin{thebibliography}{10}

\bibitem{Callegari:ECCTD99}
S.~Callegari, R.~Rovatti, and G.~Setti.
\newblock A tailed tent map chaotic circuit exploiting {S$^{\mathrm{2}}$I}
  memory elements.
\newblock In {\em Proceedings of ECCTD'99}, volume~1, Stresa, September 1999.

\bibitem{Callegari:NOLTA98}
Sergio Callegari and Riccardo Rovatti.
\newblock Sample and hold errors in the implementation of chaotic maps.
\newblock In {\em Proceedings of NOLTA'98}, volume~1, pages 199--202, Crans
  Montana, September 1998.

\bibitem{Callegari:ISCAS97}
Sergio Callegari, Gianluca Setti, and Peter~J. Langlois.
\newblock A {CMOS} {Tailed} {Tent} {Map} for the generation of uniformly
  distributed chaotic sequences.
\newblock In {\em Proceedings of IEEE ISCAS'97}, volume~2, pages 781--784, Hong
  Kong, June 1997.

\bibitem{Clarkson:NNW-1993-5}
T.~G. Clarkson, C.~K. Ng, and J.~Bean.
\newblock Review of hardware {pRAM}s.
\newblock {\em Neural Networks World}, (5):551--564, 1993.

\bibitem{Delgado:EL-1993-25}
M.~Delgado-Restituto, F.~Medeiro, and A.~Rodríguez-Vázquez.
\newblock Nonlinear, switched current {CMOS} {IC} for random signal generation.
\newblock {\em Electronics Letters}, (25):2190--2191, 1993.

\bibitem{Devaney:CDS-1989}
R.~Devaney.
\newblock {\em An introduction to Chaotic Dynamical Systems}.
\newblock Addison Wesley, 2$^{\mathrm{nd}}$ edition, 1989.

\bibitem{Hughes:ISCAS93}
J.~B. Hughes and K.~W. Moulding.
\newblock {S$^{\mathrm{2}}$I}: a two step approach to switched currents.
\newblock In {\em Proceedings of IEEE ISCAS'93}, volume~2, pages 1235--1238,
  1993.

\bibitem{Kennedy:ECCTD99}
M.~P. Kennedy.
\newblock Chaotic communications: State of the art.
\newblock In {\em Proceedings of ECCTD'99}, volume~1, Stresa, September 1999.

\bibitem{Kolumban:ECCTD99}
G.~Kolumban and B.~Frigyik.
\newblock Robust chaotic communication without synchronization.
\newblock In {\em Proceedings of ECCTD'99}, pages 445--448, Stresa, August
  1999.

\bibitem{Kolumban:TIEICE-81A-9}
G.~Kolumban, G.~Kis, Z.~Jako, and M.~P. Kennedy.
\newblock {FM-DSCK}: a robust modulation scheme for chaotic communications.
\newblock {\em IEICE Transactions on Fundamentals}, E81-A(9):1798--1802, 1998.

\bibitem{Lasota:CFN-1995}
A.~Lasota and M.~C. Mackey.
\newblock {\em Chaos, Fractals and Noise. Stochastic Aspects of Dynamics}.
\newblock Springer-Verlag, 2$^{\mathrm{nd}}$ edition, 1995.

\bibitem{Rovatti:ECCTD99}
R.~Rovatti, G.~Setti, and S.~Graffi.
\newblock Chaos based {FM} of clock signals for {EMI} reduction.
\newblock In {\em Proceedings of ECCTD'99}, volume~1, Stresa, September 1999.

\end{thebibliography}

\end{document}